# Memetics and Neural Models of Conspiracy Theories

**Włodzisław Duch**[1]

**Abstract.**
Conspiracy theories, or in general seriously distorted beliefs, are widespread. Large number of books and articles has been written on this subject from psychological, sociological and political science perspectives. How and why conspiracy theories transpire in the brain is still more a matter of speculation rather than of hard science. In this paper one plausible mechanisms is investigated: rapid freezing of high neuroplasticity (RFHN). Emotional arousal temporarily increases neuroplasticity leading to creation of new associations, pathways that spread neural activation. Using the language of neurodynamics *a meme* is defined as quasi-stable associative memory attractor state. Depending on the temporal characteristics of the incoming information and the plasticity of the network, memory may self-organize, creating memes with large attractor basins, drawing in many accidental patterns representing concepts and memories. Creation of such memes with numerous fake associations distorts relations between stable memory states, redefining world-view. Simulations of neural network models trained with competitive Hebbian learning (CHL) on stationary and non-stationary input data show the formation of distorted memory states. Short learning with high plasticity followed by rapid decrease of plasticity leads to memes with large attraction basins, distorting input pattern representations in associative memory. Such system-level models may be used to understand formation of strong attractor states of the neurodynamics, representing distorted deeply settled beliefs.

## I. Introduction

BELIEFS in conspiracy theories are a part of much wider subject: formation of beliefs, creation of memes, distorted memories, twisted worldviews, or in general investigating ways in which learning fails to represent the data faithfully. Artificial neural network community has focused on faithful learning methods, but there is another, neglected side of learning and memory formation. When observations are not learned perfectly, what types of errors one may expect, and how they influence beliefs and actions of the person or artificial system? Which observations will be neglected, which will be remembered, which will be transformed into memes that are likely to be transmitted in a distorted form to brains of other people? The world view that we use to guide our behavior is based on a network of associative memory states. Consolidation of new memory states in the neocortex may occur quite quickly if they are well connected to other memory states [1]. Several lines of research lead to this conclusion: animal studies, association of places with items in mnemotechnics, behavioral studies on the use of schemas for rapid learning and building of cognitive maps. Neural models of schemas and sequences of associations may be based on attractor states in neural networks [2]. Each episodic or semantic memory state is based on activations of synchronized, distributed network of brain regions. Using functional magnetic resonance imaging (fMRI) evoked by natural movies Huth et al. [3]have created "semantic atlas", showing distribution of brain activity for categories of hundreds of objects and actions.

[1] Włodzisław Duch, Dept. of Informatics, Faculty of Physics, Astronomy and Informatics, and Neurocognitive Laboratory, Center for Modern Interdisciplinary Technologies, Nicolaus Copernicus University, Toruń, Poland, E-mail: wduch@umk.pl

Each specific activation pattern may be represented by an attractor state of neurodynamics. It is linked in many ways to other activation patterns in the brain. Transitions between these patterns define trajectory of brain/mental states. These transitions depend on current knowledge schemas, history of previous activations (priming), general emotional state, specific context cues that invoke memories, and many other factors. Transitions that happen frequently increase probability of association between different activation patterns [5]and may not only create strong association but distort or even completely blend different memories, creating false memories [6][7]. In some cases memories may become easily activated in various contexts, leading to false associations, schemas that develop into conspiracy theories. While there is a large body of literature on conspiracy theories written by historians, philosophers, psychologist, sociologists or political scientists (Routledge has a whole series of books on conspiracy theories), our understanding of the brain mechanisms is completely lacking. The best explanations that we have relate beliefs in conspiracy theories to personality traits, mental disorders, or need for simple explanations.

**Memetics**, introduced in the 1976 book "The Selfish Gene" by Richard Dawkins [8]tried to explain cultural information transfer and persistence of certain ideas in societies. Memes may be understood as sequences or information structures that tend to replicate in a society. Despite great initial popularity of memetic ideas, and the desperate need of mathematical theories to underpin social science, theories connecting neuroscience and memetics have never been developed. *The Journal of Memetics* was discontinued in 2005 after 8 years of electronic publishing. Memetic ideas were relegated into a set of vague philosophical and psychological concepts of little interest to neuroscience. The lack of efforts to understand distortions of information transmission and memory storage in biological learning systems is certainly related to the lack of theoretical models, and to the experimental difficulties in searching for memes in brain activity. McNamara [9]has argued that neuroimaging technology may be used to trace memes in the brain and to measure how they change over time. Following Heylighen and Chielens [9][10]*memotype* and *mediotype* distinction he proposes to distinguish *i-memes*, internal activation of the central nervous system, from the external transmission/storage of information structures, the *e-memes* existing in the world (for example, created by marketing, or various media advertisements). One should distinguish clearly abstract information structure of memes, and their implementation in the brain or in artificial cognitive system. Internal representations of i-memes are created by forming memory states that link neural responses resulting from e-meme perception to behavioral (motor) responses that are necessary for replication of memes, linking sensory, memory and motor subsystems in the brain. Sets of memes forming *memeplexes* determine world views, including culture, values and religions, predisposing people to accept and propagate selected memes. Brain research has made a great progress in understanding schemas in the last decade [11]. Perhaps the time is ripe to make some progress along these lines to link the concept of memes with memory mechanisms that facilitate their spread. This could have very important social and educational implications [12].

In the fascinating book "Why people believe weird things" Michel Shermer writes about 25 fallacies that lead people to believe in conspiracy theories and other bizarre things [11]. Brains are predisposed to perceive various observed patterns as meaningful information (pareidolia), search for explanations and form theories, referring to the long-term episodic and semantic memory. Conceptual framework that is needed to interpret new observations includes memes is activated by various cues that invoke memory associations. Memes that are strongly encoded certainly influence most mental processes. Observations that agree with established individual beliefs will lead to strong activations of brain networks, thanks to the mutual co-activations of memeplex patterns, creating additional memes that make the whole memeplex even stronger. Contradicting arguments, facts or observations will arouse only transient weak activations of brain networks and will be ignored. Worse than that, mentioning or presenting anything that may retrieve memes will only increase their influence, contributing to stronger encoding and easier arousal of false associations. The levels-of-processing paradigm in memory research has found now support in neuroimaging of deep and shallow episodic memory encoding, modulated by a number of neurotransmitters and linked to emotional arousal [14]. Research on forgetting shows that retrieval of competing memory traces may lead to interference and suppression of weaker patterns [15]. If conspiracy memes are already deeply encoded they will erase memory of contradicting facts.

Science systematically tries to falsify hypothesis by performing experiments, but from the evolutionary perspective falsification is simply too dangerous. In slowly changing environment stability of beliefs is more important, even at the price of wide acceptance of meaningless taboos and superstitions. Even today educational systems in most countries do not encourage skeptical thinking. Religious leaders and conservative politicians are strongly opposing instating skepticism into the educational system, in fear of destabilization of established world views. There is little or no penalty for accepting false beliefs by individuals. Mutual support within groups of believers gives boost to distorted views of reality, leading to bizarre conspiracy theories.

Discussion presented above shows that fake news and conspiracy theories tap into basic brain mechanisms of memory and learning. The complexity of the belief formation processes has discouraged scientists from approaching this important problem. Obviously no simple computational model is going to explain all facts related to formation and preservation of human beliefs, and in particular of conspiracy theories. This should not discourage us from forming testable hypothesis based on neurodynamics. After all simple neural network models introduced by Hopfield and Kohonen, despite being only loosely inspired by neurobiology, have found a number of applications in computational psychology and psychiatry [16], and the central role of large scale neural dynamics as a basis for understanding brain processes is now well recognized [17][18]. The two main goals of this paper are thus to show that memetics may be based on solid theoretical foundations grounded in neurodynamical models, and that learning using simple memory models may help to understand the process of formation of conspiracy theories. Although only simple competitive learning models are used in this paper it should open the road towards application of more complex neural models that link memetics with neuroscience.

The next section introduces memetics and discusses representation of information in the brain. It is followed by a section on competitive learning models of memory formation. These models are used to illustrate some mechanisms of memory distortions. Remarks about implications of network simulations for the theory of memetics are presented in section four, and the final conclusions are in section five.

## II. Memetics and information in the brain

### *A. Subjective information*

Ultimately all thoughts and beliefs result from neurodynamics. The flow of neural activation through neural systems is determined by many biological factors, including brain connectivity, concentration of neurotransmitters, emotional arousal, priming effects, brain stem activity. Information is acquired and internalized in the brain through direct observation of patterns in the world, including communication with people and animals, and indirectly through various media, texts and physical symbols of all sorts. Brains provide material support for mental processes, understanding and remembering symbols, ideas, stories. Memes are units of information that spread in cultural environment, information granules that prompt activation of patterns in brains molded by particular subculture. Therefore the same information may become a meme in some brains, and may be ignored by other brains.

Understanding is a process that requires association of new information with what has already been learned. New things are learned on the basis of what is already known by the system. This is a general principle behind brain activity, information gain should be measured as a change induced in cognitive systems [19]. Patterns are encoded in memory depending on the context, sequence of events, attention devoted to these patterns, association with known facts, properties of already encoded information, and general mental state during the encoding process. Definition of Shannon information as entropy does not capture the intuitive meaning of the value of information for the cognitive system. The amount of optimal restructuring of the internal model of the environment (optimal in the minimum length description sense [20]) resulting from new observation (i.e. a new meme added to the memeplex) is a good subjective measure of the quantity of meaningful information carried out by this observation. **Pragmatic information** that captures the subjective meaning of information is based on the difference between algorithmic information before and after observation is made [19][19]. Itti and Baldi used similar idea to define the amount of surprise, measured as the relative entropy or Kullback-Leibler (KL) divergence between the posterior and prior distributions of beliefs in Bayesian models [21].

### *B. Memes in brains*

In memetics information structures that reflect part of mental content based on a network of memes are called memeplexes. They evolve in response to enculturation and exposure to observed patterns. Specific cultural behaviors, learned concepts, word meanings, collocations or phrases describing ideas, may be treated as memes. Some are very rare and difficult to acquire, while others spread quickly with ease. Mental content can be much wider than just the network of memes. Memetics should position itself in respect to the theory of communication, language acquisition and learning.

Using the language of neurodynamics **a meme** is defined as a **quasi-stable associative memory attractor state, with robust attractor basin**. Brain activation $A(w)$ prompted by stimulus $w$ (a word, set of words, seeing a symbol) may rapidly evoke activation corresponding to meme $A(w) \rightarrow A(M(w))$. The same attractor state may be activated by many different stimuli, including purely internal activations. For simple visual percepts, such as shapes of objects, similarity between brain activations $A(M)$ in the inferotemporal cortical area have been directly compared, using fMRI neuroimaging, to the similarity of the shapes of these objects [22]. Significant similarity has also been found in the fMRI patterns of whole brain activity when people think about specific objects [3][4][23], showing how meaning of concepts is encoded in distributed activity of the brain. Such encoding may be used for brain-based vector representation of the semantic meaning in natural language processing algorithms [24]. Similarity between memes corresponding to perceived objects $M_i \Leftrightarrow O_i$, may be roughly compared to some measures of similarity between object properties. Therefore similarity between brain activities $A(M_1)$ and $A(M_2)$ that represent two memes $M_1$ and $M_2$ evoked by objects $O_1$,$O_2$ (percepts, cues, words) should be directly related to some measures of object similarity:

$$S_a(A(M_1),A(M_2)) \sim S_o(O_1,O_2). \quad (1)$$

McNamara [9] hopes to detect the signature patterns of new memes by analyzing the neurodynamics of learning novel name–action associations for abstract category names, looking at the changes of the brain connectivity profiles. This may be a useful strategy for abstract categories, or for simple percepts, but general search for signatures of memes using neuroimaging techniques will be very difficult. Activation patterns may significantly differ for individual people, depending on their memeplexes. For the same person distribution of fMRI activations may change at different times of the day. Transcranial magnetic stimulation (TMS) disrupting the function of the left inferior frontal gyrus has already been used to alter belief formation in favor of remembering more bad news [25]. Such brain stimulation may be used to change acceptance of memes that would normally be ignored.

Many concepts gradually change their meaning over time. For example, the concept of a gene has significantly changed in recent years. Genes, once defined as sequences of DNA base pairs that code proteins, are now understood as distributed DNA and RNA templates, with exons on different chromosomes, "encoding a coherent set of potentially overlapping functional products" [26]. Precise definition of a gene is difficult because they are structures of partially mutable highly organized molecular matter living in specific network of complex processes. They exist because highly specialized environment facilitates their replication. Strong coupling of all elements in this environment makes the concept of a gene rather fuzzy: it is not a simple DNA sequence, but a complex pattern in the whole network of processes, active only in certain situations controlled by epigenetic factors. The whole system is responsible for replication of information. Memes are even more difficult to extract from the whole network of brain activities, they exist more as vortices in neurodynamics that actualize only in certain context than as separate entities.

Understanding how brain connectivity and other factors determine neurodynamics, encode beliefs, filter incoming information, distort it and transmits it further, is certainly a grand challenge. Complex information processing in the brain has not yet been understood in sufficient details to allow for development of comprehensive theories of such processes, but some insights based on simple memory models may be gained. New information added to the memeplex (existing pool of interacting memes, or attractor states) becomes distorted, changes the memeplex, and is replicated further. Once a set of distorted memory states is entrenched it becomes a powerful force, attracting and distorting all information that has some associations with these states, creating even broader basins of attractors. Encoding of information in this way enhances the memeplex and is one of the reasons why conspiracy theories are so persistent.

### *C. Concepts in brains and in computers*

In the Natural Language Processing (NLP) field word meaning is approximated using correlations between co-occurrence with several adjacent words. Vectors storing these correlation coefficients $C(w)$ represent words $w$ by averaging over many contexts restricted to a specific meaning of a given word (this requires annotation of large text corpora). From the human point of view faithful representation of word meaning should require similar ordering of distances $D(C(w_1),C(w_2))$ between vectors $C(w_1),C(w_2)$ representing words $w_1$, $w_2$, as shown by dissimilarities between brain activations when concepts associated with these words are invoked:

$$S_a(A(w_1),A(w_2)) \sim D(C(w_1),C(w_2)) \qquad (2)$$

Each vector $C(w)$ attempts to approximate meaning of the word that is encoded in the distribution of brain activity [23][24]. Without priming effects [27] and association of words with existing memory patterns only a very coarse representation is possible. Brain activations strongly depend on context, and therefore the distance function $D(C(w_1),C(w_2);\text{cont})$ should be context dependent. The whole process is dynamic, with spreading of neural activations responsible for priming related concepts and providing feedback that becomes part of the new pattern encoding. Meaning is thus connected to the activation of many subnetworks in the brain, memory of sensory qualities and motor affordances. Dynamical approach to the NLP vector model has not yet been fully developed although some steps in this direction have been made [28][29]. Despite our efforts (Duch, unpublished) to describe dog breeds in terms of skin, head and body features derived from databases and semi-structured texts describing dogs, it was not possible to categorize accurately dog breeds only by their features. Using images (or just silhouettes) of dogs leads to more accurate and faster identification of dog breeds. Brain activity evoked by hearing or reading words evokes internal imagery at a high level of invariant, multimodal object recognition. Similarity functions between objects $S_o(O_1,O_2)$ based only on correlations between verbal descriptors, cannot do justice to estimations of similarity of brain activations. Finer discrimination may require recall of lower-level sensory qualities, referring to particular shapes, colors, movements, voice timbre or tastes. Vector representation based on word correlations does not reflect essential properties of the percep-tion-action-naming activity of the brain [30], it does not even contain structural description in terms of object features or phonology. More details on word representation in the brain and its relation to the vector model may be found in [24][28]. Words are only labels that point to internalized knowledge. Representation of percepts arising from sensory imagery is a minimal requirement for NLP systems capable of semantic interpretation of concepts.

In the next section competitive learning models are introduced, and then used to illustrate the process of learning that leads to memes based on distorted relations.

## III. Competitive learning and weird beliefs

Conspiracy theories have serious consequences for politics, especially environmental policies and health, they facilitate growth of political extremists and dangerous religious sects [13][30]. Conspiracy theories are investigated mainly by sociologists and psychologists, focusing on hidden networks controlling political and economic factors that are poorly understood. Instead of analyzing why and how brains form weird distorted views of reality, they invent vague concepts and construct theories that are impossible to connect with brain research. While there are many psychological reasons for formation of such beliefs, so far there have been no attempts to create a cognitive theory supported by computational models, capable of generating testable hypotheses. In the past secret societies were rather rare, but now media try to stir controversy discussing GMO, vaccines, AIDS, miracle cures, UFOs, prophecies, assassinations, airplane crushes and other such issues, despite plausible explanations based on scientific arguments or on common sense consensus.

The language of memetics is descriptive and does not help to explain deeper reasons why some information become memes and others are forgotten [8][32]. Conspiracy theory may be treated as a memeplex that is easily activated by various pieces of information, giving it meaning consistent with the memeplex responses. From neurobiological perspective learning requires adaptation, changing functional connectivity, adjusting physical structure of the brain. Learning is thus energy-consuming, requires effort that should be carried out only when there are potential benefits. Simple explanation of complex phenomena have thus a great advantage even when they are quite naïve, as long as they do not lead to behaviors that are obviously harmful, or significantly decrease chances for reproduction. Evolutionary Darwinian adaptations are established only after several generations, and have noticeable influence on human beliefs only if they affect large subpopulations. Evolutionary factors explain slow changes in approaches to human freedom, caste and racial divisions, abandonment of slavery, attitudes towards children (selling children into slavery continued until 19$^{th}$ century) etc. Weird beliefs have more plausible explanation in distortions of the rapid learning process. However, the field of neural networks aiming at achieving perfection in learning paid little attention to distortions of learning.

There are many scenarios that may lead to formation of distorted views of observations. Slow and steady environmental pressures lead to changes of attitude and may redefine the whole world view. Here I will focus on a rather common situation that arises as the result of rapid decrease in neuroplasticity. Emotional arousal coming from uncertainty of important information (ex. rumors that something potentially life-threatening has happened) leads to confusion and strong anxiety (the rumors may not be true, it is not clear what has really happened). High emotions and stress are linked to release of large amounts of neurotransmitters and neuromodulators from the brain stem nuclei, through the ascending pathways, activating serotonin, norepinephrine, acetylcholine and dopamine systems. Strong arousal increases brain plasticity facilitating rapid learning of all potentially relevant cues. As it is not yet clear what in the end will appear as important piece of information all facts and gossips should be memorized. Uncertainty may persist for longer time, but after some period the situation may become clear, strong arousal will subside, sources of neurotransmitters will be depleted and neuroplasticity will rapidly decrease. Thus the best recipe for reality distortion is strong and rather persistent emotional arousal, uncertainty of information, gossip and suspicions, followed by a tragic end leading to depression. The system is left with memories of gossips frozen in its associative memory. All future information related to the event will be associated and interpreted in view of what has been memorized at that period, setting foundations for conspiracy theory.

This scenario may be reproduced in many unsupervised competitive learning neural models [16], including ART model that has vigilance parameter [33] to regulate neuroplasticity. Many other competitive learning models based on Hebbian learning have been presented in [34]. DemoGNG 2.2 Java package, written by Bernd Fritzke and Hartmut S. Loos [35], implements winner-take-all learning in Self-Organizing Map (SOM), Competitive Hebbian and Hard Competitive Learning, Neural Gas, Growing Neural Gas, Growing Grid, and other algorithms [36]. In all these algorithms activity of units representing neurons is compared with the input, and those units with the best match adapt their parameters increasing their activation. Neurons in the neighborhood of a winner are also allowed to adapt, depending on their distance from the winner. If there is no clear match constructive algorithms add new neurons allowing the network to grow.

The rapid freezing of high neuroplasticity (RFHN) model described here is based on the following assumptions:

- Emotions and uncertain stressful situations at the beginning of learning lead to high neuroplasticity.
- High neuroplasticity is imitated in the model by large learning rates (due to the primary neurotransmitters), and by a broad neighborhood of the winner neuron for each input pattern (due to the diffuse neuromodulation and volume learning).
- The network tries to reflect associations between input vectors, adapting neuron parameters (usually codebook vectors) to approximate distribution of information contained in the presented input vectors.
- Sudden decrease of the uncertainty and emotional arousal is mirrored by the decrease of learning rates and neighborhood sizes, leading to distortions of complex relations between input items.
- Slow forgetting that follows rapid freezing is based on memory reactivations, and contributes to the retention of memory states represented by the highest number of neurons only, forming clusters of nodes with large and strong basins of attraction that link many states.
- Clusters of neurons that are frequently activated and thus easily replicated represent memes.
- Conspiracy theories are characterized by memplexes, numerous strong memes, with many neurons encoding information that has never been presented, forming distorted associations between facts.

As a result these networks do not reflect real observations. The role of emotions in susceptibility to fake news has been verified in a recent experiment [37]. The RFHN model may be simulated using several competitive learning models. In fact all such models show similar behavior, therefore only the Self-Organized Maps [16], and Neural Gas model with Competitive Hebbian Learning (NG-CHL) [35] are shown below for illustration.

The basic idea of competitive learning is to approximate the activity of neural cell assemblies by neurons (units) that serve as codebook vectors $\mathbf{W}(t)$. They represent receptive fields, adapting to the probability density of the incoming signals. Each neuron receives input signals and competes with other neurons using the winner-takes-most (or takes all) principle, leaving only a small subset of active units that are updated. The winning neural assembly is represented by a vector $\mathbf{W}^{(c)}(t)$ and a small group of vectors in its direct neighborhood $O(c)$. SOM starts with a fixed two-dimensional grid of neurons. Learning proceeds by identifying most similar codebook vector to the current observation $\mathbf{X}(t)$, and updating the codebook vector and vectors in its immediate physical neighborhood according to the formula:

$$\begin{aligned} &\text{For } \forall i \in O(c) \\ &\mathbf{W}^{(i)}(t+1) = \mathbf{W}^{(i)}(t) + h(r_i, r_c, t)\left[\mathbf{X}(t) - \mathbf{W}^{(i)}(t)\right] \end{aligned} \tag{3}$$

where the neighborhood is usually assumed to be Gaussian:

$$h(r, r_c, t, \varepsilon, \sigma) = \varepsilon(t) \exp\left(-\left\| r - r_c \right\|^2 / \sigma^2(t)\right) \tag{4}$$

The size of this neighborhood is decreased from the initial value of dispersion $\sigma_i$ to the final value $\sigma_f$ according to the formula:

$$\sigma(t) = \sigma_i \left( \frac{\sigma_f}{\sigma_i} \right)^{t/t_{\max}} \tag{5}$$

The maximal age $t_{max}$ determines the annealing schedule. The learning rate is similarly decreased by:

$$\varepsilon(t) = \varepsilon_i \left( \frac{\varepsilon_f}{\varepsilon_i} \right)^{t/t_{\max}} \tag{6}$$

SOM model has been used with success in many applications, for example it works quite well, in comparison to other neural models, for explanation of details of orientation and ocular dominance columns in the visual cortex [38].

The NG-CHL algorithm does not have such fixed initial grid topology as does SOM, new neurons are recruited for encoding input patterns like gas particles. At each adaptation step a connection between the winner and the second-nearest unit is created, if it does not already exist. The newly created or existing selected edges are refreshed receiving age=0, while the age of other edges emanating from the winner neurons are increased by 1. The reference age is gradually changed from $T_i$ to $T_f$ according to:

$$T(t) = T_i \left( \frac{T_f}{T_i} \right)^{t/t_{\max}} \tag{7}$$

Edges that are not refreshed for more than $T(t)$ steps are removed. This simulates forgetting mechanism.

Following computational experiments have been done to illustrate RFHN model:

- Training SOM and NG-CHL on stationary data concentrated in two distinct areas, with initial high plasticity and rapidly decreasing learning rates.
- Training SOM and NG-CHL on non-stationary data from observations that move and suddenly change, with initial high plasticity and rapidly decreasing learning rates.
- Retraining the model after malformation of relations has already occurred, using temporally increased plasticity.

The number of neurons in the brain is extremely large, therefore it is instructive to check how the number of network nodes in simulations will affect distributions. For the stationary experiments 10.000 nodes have been used, with initial parameters randomly distributed, and signals coming from two separated circular areas. This should represent two alternative situations that are monitored. For the non-stationary situation all parameters were initially concentrated in the rectangular patch, simulating situations in which restricted domain has already been learned and stable. Then the patch moves across the whole domain providing new input patterns (observations) from the areas it covers. When the edge of the domain is reached the patch jumps to the other side.

## IV. Conspiracies and memory distortions

### *A. Stationary situation*

Perfect representation of all signals should cover two distinct circular areas. A good solution that requires slow learning with 500.000 steps is shown below. The domain and relations (represented by edges) of input patterns are represented fairly well.

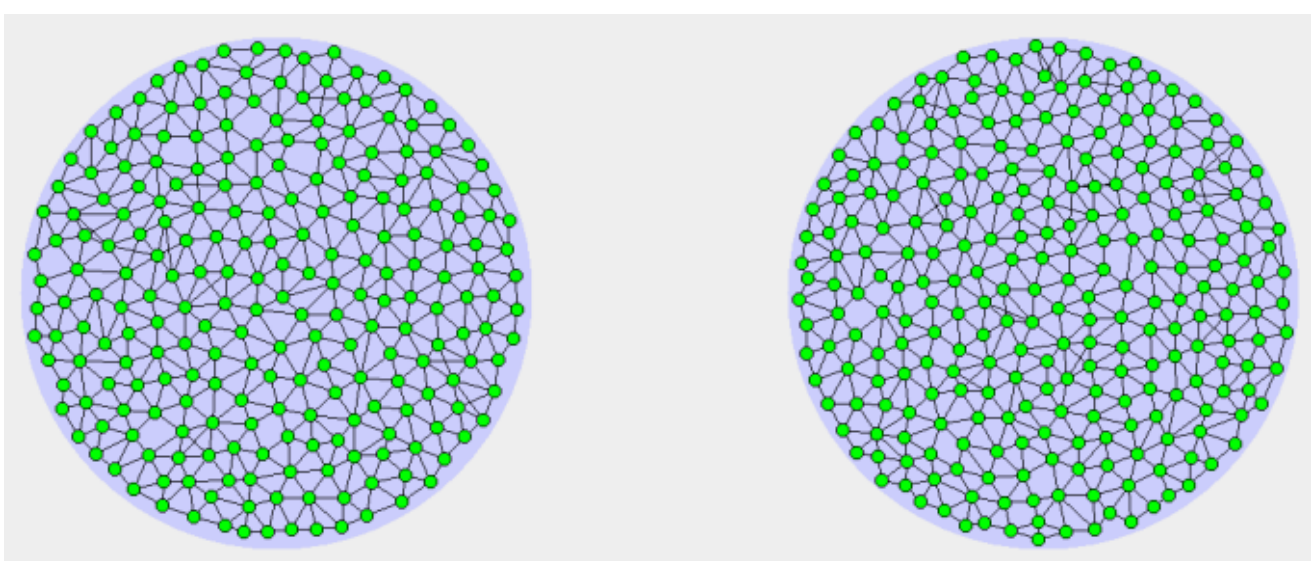

Fig 1. SOM network learning slowly stationary uniform samples drawn from double circles approximates these distributions correctly.

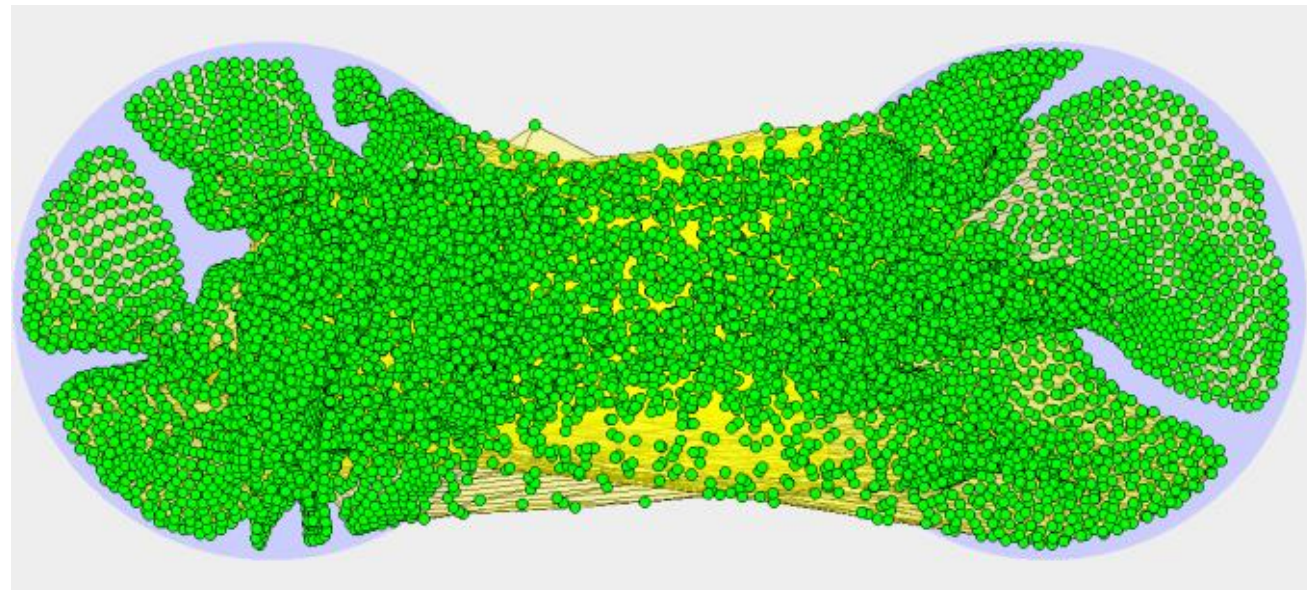

Fig. 2. SOM network learning the same distribution as in Fig. 1, with fast decrease of plasticity covers areas where no samples appeared and leaves large gaps in the data space.

Training 100x100 SOM network, with initial $\sigma_i$=5, $\sigma_f$ =0.01, $\varepsilon_i$=1, $\varepsilon_f$ =0.001, for 10.000 steps, did not pull all parameters of neurons towards data area. Despite high density of neurons some gaps have been left and were not removed by further retraining. This effect comes from the dynamics of learning with shrinking neighborhoods. There is a greater chance for neurons near the edge to be pulled towards high density areas by many neurons that are selected as winners than to be pulled towards the data in the gap area. Moreover, in the space where no samples ever appeared many neurons are placed, and this will lead to false associations and confabulations (Fig. 2). These effects are random due to stochastic nature of learning. The resulting map has the same character although details differ every time it is simulated.

The NG-CHL model with initial high plasticity and rapidly decreasing learning rates has also produced big gaps and high density areas, as seen in Fig. 3. Forgetting parameters have been set to $edge_i$=20 and $edge_f$ =200. Further retraining with fast forgetting creates even bigger gaps. Many input patterns are therefore associated with high density clusters acting as memes. Associations with other input patterns are based more on stereotypes (clusters) rather than faithful observations.

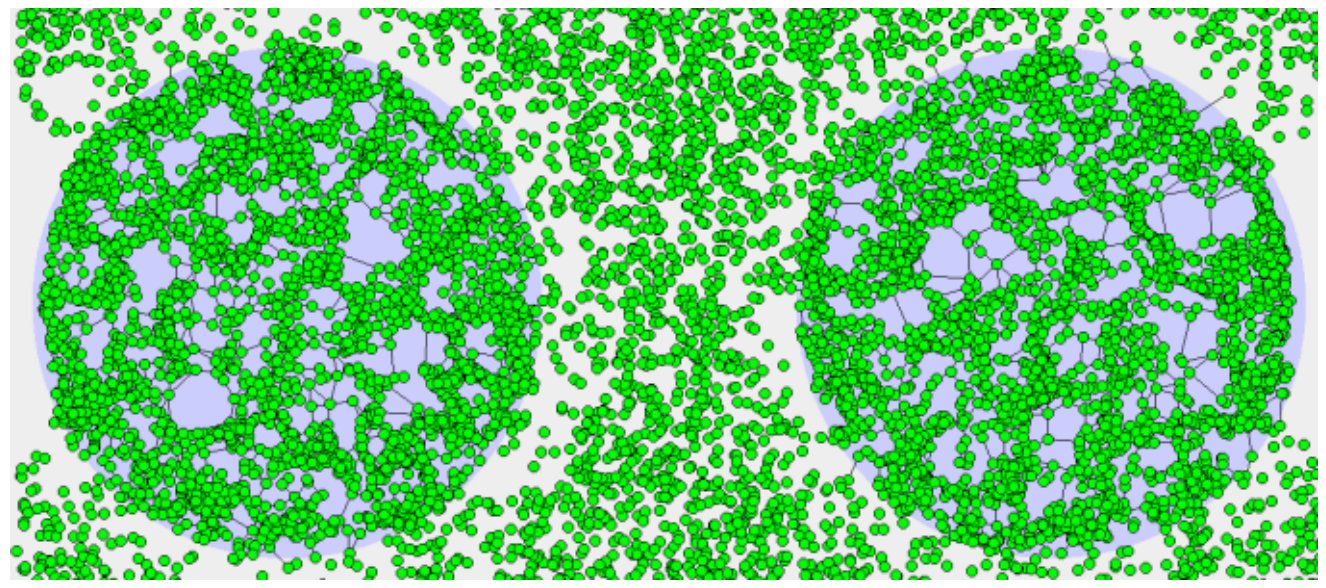

Fig. 3. The Neural Gas model with fast decrease of plasticity creates even stronger distortion of original distribution than SOM map in Fig. 2, leaving many gaps and covering empty space densely.

### *B. Non-stationary situations*

Learning in nonstationary situation is much more difficult and therefore distortions are much stronger. Using the same parameters as for the stationary case SOM started with high plasticity and that was rapidly decreased (in 10.000 steps). The map in Fig. 4 shows very strong concentration of neurons that point to the initial patterns, and the network did not learn much during later part of the training. It has ignored most of the facts coming after rapid learning period, creating one big sink for all associations. Such network will interpret most input data as similar to what it has seen in the critical period of high plasticity.

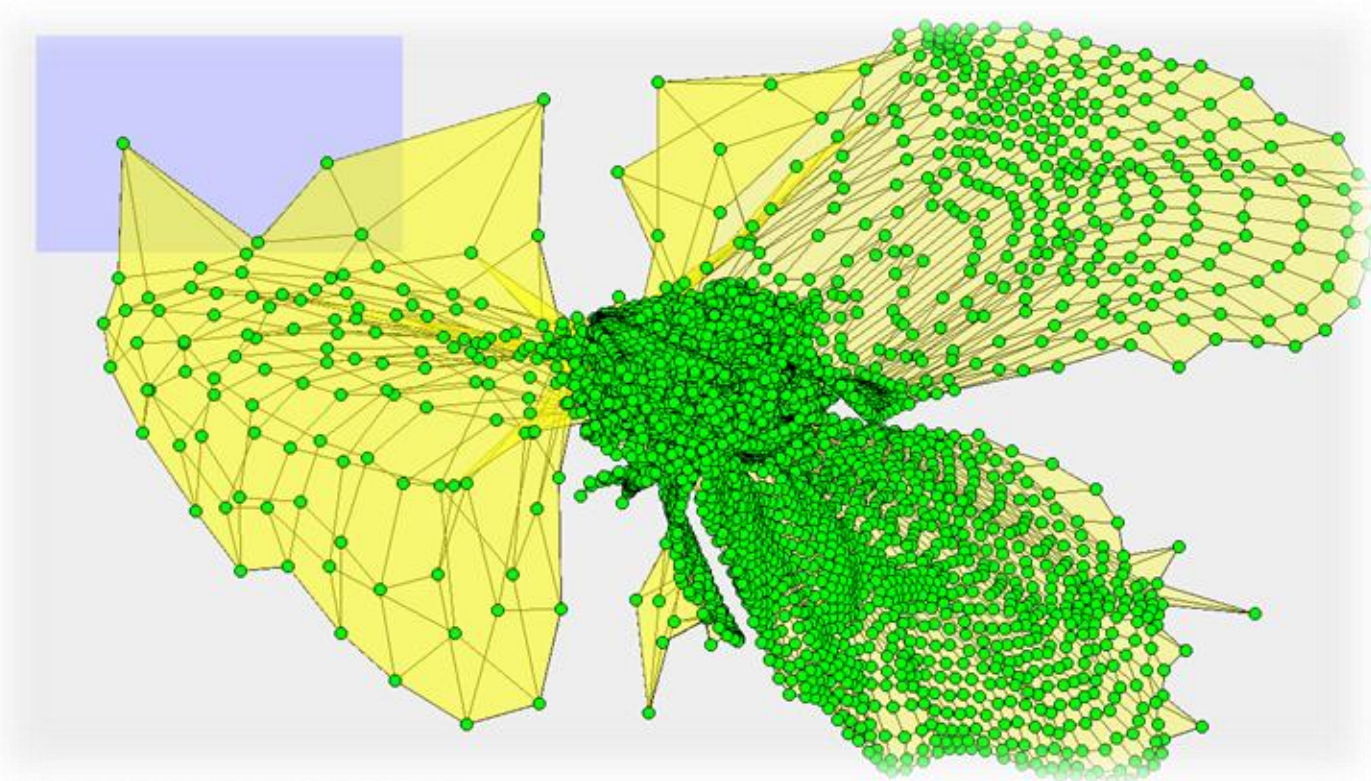

Fig. 4. SOM with rapidly decreasing plasticity for non-stationary distribution; samples come here from a moving square (seen in the left corner) and with very slow learning are uniformly distributed in the whole rectangle, but fast learning leads to completely.

Further training with increased plasticity may somehow repair the distorted view, although even after a very long training (Fig. 5) strong meme that has been formed in the center is still present. Large basin of attraction for this meme will lead to its activation frequent activation even by irrelevant input patterns. After additional 100.000 steps the central sink has loosened providing still distorted, but more diversified map.

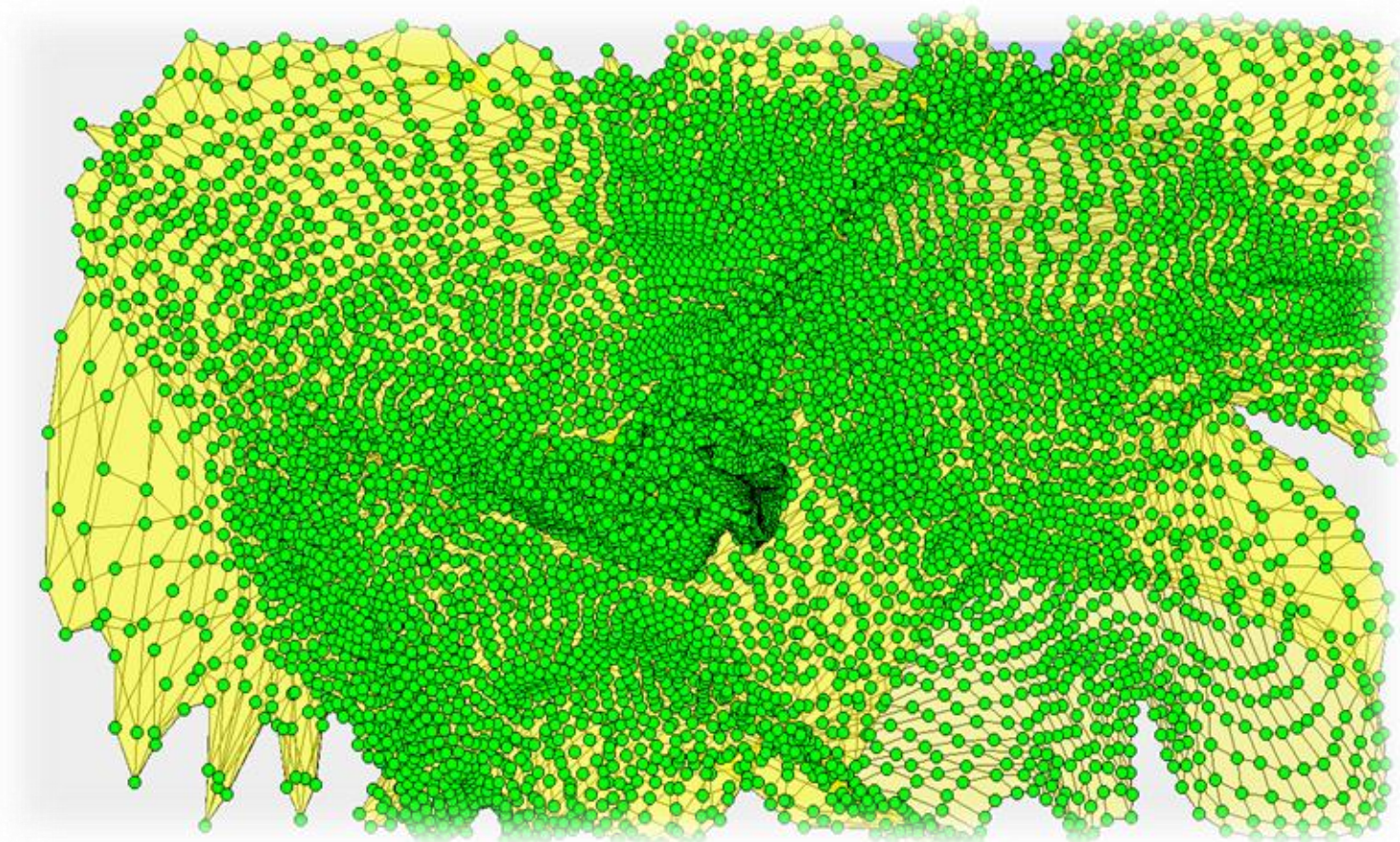

Fig. 5. Non-stationary case, Neural Gas map as in Fig. 4, followed by long slow training (100.000 steps) only partially recovers uniform distribution, leaving large concentration of the codebook vectors in the middle.

The NG-CHL algorithm may also create completely distorted representation. After 40.000 steps with rapid decrease of plasticity it has created two separate memplexes, each with several strong memes that are used to interpret all incoming patterns (Fig. 6).

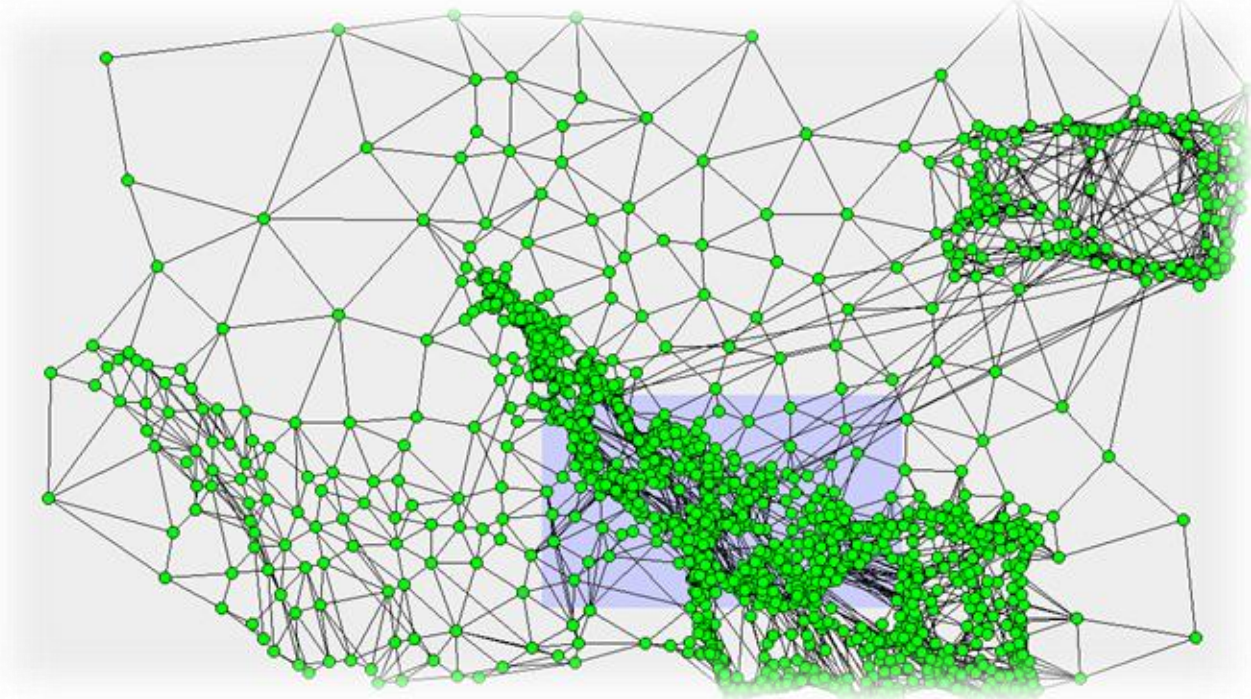

Fig. 6. In the non-stationary case Neural Gas created two densely connected structures.

Maps created with rapid decrease of initially high plasticity are quite unstable. In Fig. 7 another solution is shown with 4 larger memeplexes that completely distort view of the input patterns. It is quite difficult to create faithful representation of input patterns for non-stationary signals. Very long training times with several hundred thousand iterations are needed to achieve it.

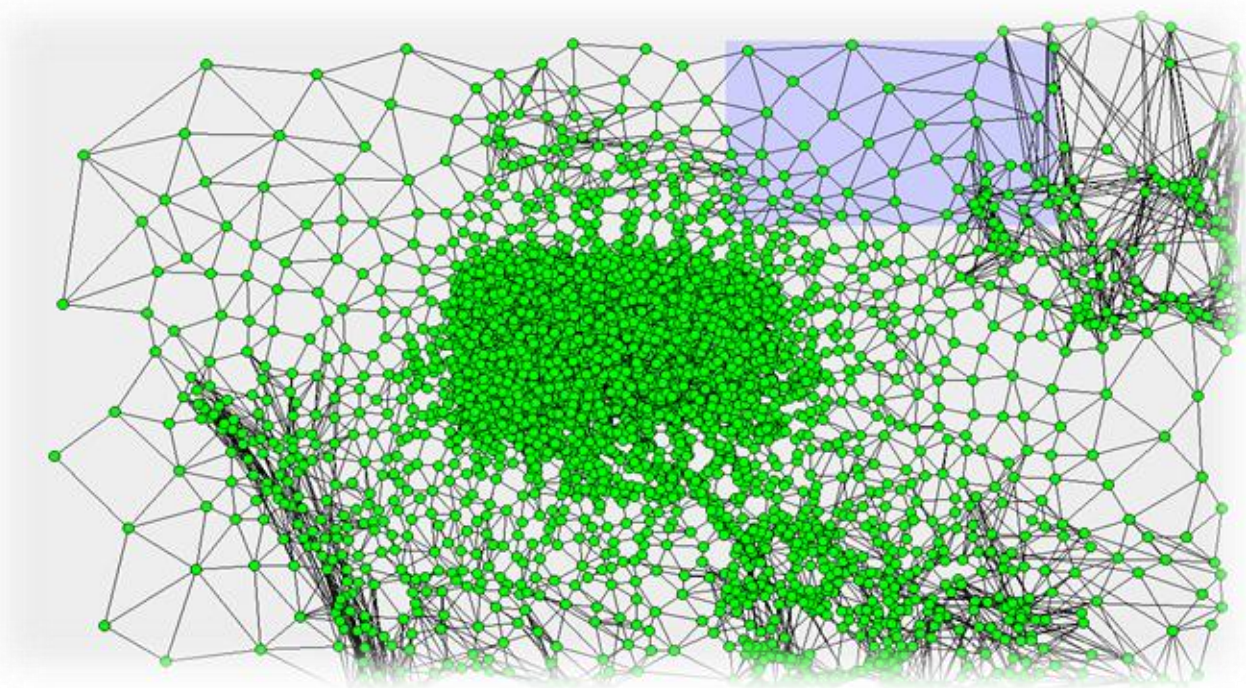

**Fig. 7. Another neural gas map for non-stationary case, showing how unstable such learning may be.**

In case of rapidly changing situations it is much more likely that a distorted view will be learned instead of a faithful representation of reality. Gaining experience in changing environment obviously takes more time, as can be observed in many domains such as medicine, where initial background knowledge is slowly structured into high competence by working environment.

## V. Conclusion

Biological and psychological belief forming mechanisms are very complicated. Predispositions for accepting distorted views of reality may come as a side effect of education and life experiences and therefore are rather hard to investigate. Accepting simple explanations is rewarding, creates pleasant feelings of understanding. Complex explanations requires a lot of effort and a long time to understand them fully. A simple (although inadequate) explanation is always better than to have no explanation at all, saving energy required for learning and creating a (false) impression of reducing uncertainty.

Why do people believe in conspiracy theories? Because this is how their brains work. Neurodynamics helps to understand the conditions under which large basins of attractions, called memes, are created in memory networks, how and why they form memplexes that lead to the distorted associations. This is an important step towards linking memetics with theoretical and experimental brain science. Perhaps patterns of brain signals corresponding to memes can be measured [9], and computer simulations should help to define most suitable experimental conditions. With the advent of highly detailed brain simulations and neuroimaging techniques we should be able to understand precisely the mechanism behind false memory formation. However, it should be possible to repeat the experiments on artificial distributions with maps based on texts in some restricted domain. Each network node will represent than a word and distances between words will be based on their similarity in a given context. Such models should allow for semi-realistic analysis of formation of distorted world views.

What lessons can we draw from computational experiments with competitive learning? The rapid freezing of high neuroplasticity (RFHN) model presented here is very simple, but it seems that all types of competitive learning models show similar behavior. More complex models with high-dimensional input patterns almost certainly will have even bigger problems with faithful representation of input patterns using the rapid freezing of neuroplasticity scenario, and will lead to large attractor basins that can be interpreted as memes. Slow learning leads to faithful representations, but if the information is false (for example, frequently repeated in media) it may also end in conspiracy theory. A lie repeated ten thousand times becomes truth, as the famous "Big Lie" propaganda technique.

The contributions of this paper are two-fold. First, memetics theory has been developed in social sciences but a link to neuroscience has been missing. Linking memes to attractors of neurodynamics should help to give memetics solid foundations. Second, analysis of formation of weird beliefs is very important, but so far there have been no models of brain processes that could explain creation of such beliefs. Simulations presented here should draw attention to the need of analysis of the type of distortions that are common in neural networks. Of course more complex neural models will be needed to allow for predictions that could be compared with the results of neuroimaging and behavioral experiments, but even such coarse models based on competitive learning networks may serve as an illustration of putative processes responsible for formation of various conspiracy theories. Our next step is to perform such simulations on real data from the newspapers. Other computational models that deal with false memories, such as ART [33] and Associative Self-Organizing Network (ASON) that have been used to explain emergence of false memories [39] and can be used also to model conspiracy theories should also be considered.

## Acknowledgment

This work has started when I worked as Nanyang Visiting Professor in the School of Computer Engineering, Nanyang Technological University, Singapore, and is now supported by the Polish National Science Foundation grant UMO-2016/20/W/NZ4/00354.